\documentstyle[11pt]{article}
\makeatletter
\def\section{\@startsection{section}{1}{\z@}{-3.5ex plus -1ex minus -2.ex}
{2.3ex plus .2ex}{\Large\bf}}

\def\subsection{\@startsection{subsection}{2}{\z@}{-3.25ex plus
 -1ex minus -2.ex}
{1.5ex plus .2ex}{\bf}}

\def\vsn{\vskip 1pc \noindent}
\def\f{\newline}
\def\e{\varepsilon}

\def\comp{{\rm comp}}

\def\rr{{\bf R}}

\def\P{{\rm{\bf P}}}
\def\E{{\rm{\bf E}}}

\newcommand{\be} {\begin{equation}}
\newcommand{\ee} {\end{equation}}
\newcommand{\bd} {\begin{displaymath}}
\newcommand{\ed} {\end{displaymath}}

\newcommand{\bq}{\begin{eqnarray}}
\newcommand{\eq}{\end{eqnarray}}
\newcommand{\bqn}{\begin{eqnarray*}}
\newcommand{\eqn}{\end{eqnarray*}}
\newcommand{\ba}[1]{\begin{array}{#1}}
\newcommand{\eqa}{\end{array}}
\def\qed{
   \\[-4ex]
  \hbox to \hsize{\hfill \vrule height 1.6ex width 1.5ex
  depth -.1ex}}

\begin{document}

\bibliographystyle{alpha}

\begin{center} {\Large {\bf
RANDOMIZED AND QUANTUM ALGORITHMS 
YIELD A SPEED-UP FOR INITIAL-VALUE~PROBLEMS 
 }\footnotemark[1] }
 
\end{center}
\footnotetext[1]{ ~\noindent This research was partly supported by  
  AGH grant No. 10.420.03 \vsn}               
  
\medskip
\begin{center}
{\large {\bf Boles\l aw Kacewicz \footnotemark[2] }}
\end{center}
\footnotetext[2]{ 
\begin{minipage}[t]{16cm} 
 \noindent
{\it Department of Applied Mathematics, AGH University of Science 
and Technology,\\
\noindent  Al. Mickiewicza 30, paw. A3/A4, III p., 
pok. 301,\\
 30-059 Cracow, Poland 
\newline
 kacewicz@uci.agh.edu.pl, tel. +48(12)617 3996, fax +48(12)617 3165 }  
\end{minipage} }

\thispagestyle{empty}

\begin{center} {\bf{\Large Abstract}} \end{center}
{\small Quantum algorithms and complexity have recently been studied not
only for discrete, but also for some numerical problems. Most attention 
has been paid so far to the integration problem, for which a speed-up 
is shown by quantum computers with respect to deterministic 
and randomized algorithms on a classical computer. 
In this paper we deal with the randomized and quantum complexity 
of initial-value problems.
For this nonlinear problem, we show that both randomized and quantum 
algorithms 
yield a speed-up over deterministic algorithms. Upper bounds on the 
complexity in the randomized and quantum setting
are shown by constructing algorithms with a suitable cost, where the 
construction is based on integral information. Lower bounds 
result from the respective bounds for the integration problem.
}
\newpage

\setcounter{page}{1}
\normalsize
\pagestyle{plain}
{\Large \section{ Introduction }}
\noindent
Potential advantages of quantum computing over deterministic
or classical randomized algorithms  have been extensively studied 
by many authors 
for discrete problems, starting from Shor's paper on factorization 
of integers \cite{Shor} and Grover's algorithm for searching databases 
\cite{Grover}. 
Recently, a progress has also been achieved in 
quantum solution of numerical problems. 
The first paper dealing with the quantum complexity of a continuous problem
 was the work of Novak \cite{Novak}, who established matching upper and 
 lower bounds on the quantum complexity of 
integration of functions from  H\"older classes, based on 
the results on complexity of summation of real numbers from \cite{Bras}  
 and \cite{Nayak}.
 A general model of quantum computing
for continuous problems has been developed by Heinrich \cite{Heinrich},
where the computation of a sum of real numbers is studied
under various assumptions, and the results are applied 
to the integration problem. 
Another integration problem, computing path integrals, has been 
discussed in~\cite{Wozn}. Recently, the approximation problem in discrete 
and continuous versions has also been treated in \cite{Heinrich1} 
and \cite {Heinrich2}. The linear problems of integration and approximation 
seem to be the only specific (and important) numerical problems discussed 
in the quantum setting so far.
\f
In the randomized setting, complexity results for problems such as
integration, approximation or optimization are classical, see, e.g.,
\cite{N88} for an overview. 
\f
In this paper we deal with the randomized and quantum 
solution of initial-value problems. 
The complexity of this nonlinear problem was studied  until now 
in the deterministic worst-case and asymptotic 
 settings, see, e.g., \cite{Kac1} and \cite{Kac2} for 
 matching upper and lower bounds, or \cite{nonadapt}
 for a discussion of complexity of initial-value problems on 
 parallel computers. 
 \f
We show in this paper that  a speed-up is achieved 
for initial-value problems by randomized and quantum algorithms over 
the deterministic ones. 
The results are summarized in Theorem 1, where we establish upper 
and lower bounds on the complexity of initial-value problems 
in the randomized and quantum settings. 
The upper bounds are obtained by defining algorithms 
based on the deterministic integral algorithm developed in \cite{Kac1}. 
The procedure is shown allowing for an application of 
any algorithm for computing integrals (in the 
deterministic, randomized or quantum setting) to yield a
new algorithm for initial-value problems in the respective setting. 
In the complexity analysis, the results 
on integration from \cite{N88} and \cite{Novak} in the randomized 
and quantum settings are exploited.  
The comparison of the upper bounds on the randomized and quantum complexity to
the worst-case complexity of initial-value problems shows
that a speed-up is achieved in both non-deterministic settings.
Lower bounds on the complexity refer to those for 
integration; the existing gap between the upper 
and lower bounds is discussed.
\vsn
{\Large \section{ Problem Formulation and Results}}
\vsn
We consider the solution of a system of ordinary differential equations with
initial conditions
\be
z'(t)=f(z(t)), \;\;\; t\in [a,b],\;\;\;\;\; z(a)=\eta,
\label{1}
\ee
where $f:\rr^d \to \rr^d$, $z:[a,b]\to \rr^d$ and $\eta\in \rr^d\, $ 
($f(\eta)\ne 0$). 
\f
Given an integer $r\geq 0$, $\rho\in (0,1]$, and 
positive numbers $D_0,D_1,\ldots, D_r$ and $H$, 
we assume that the right-hand function $f=[f^1,\ldots,f^d]^T$ 
belongs to the H\"older class 
$$
F^{r,\rho} =\{\, f:\rr^d\to \rr^d \mid \; f\in C^r(\rr^d), \;\;\;
|\partial ^i f^j(y)| \leq D_i, \; i=0,1,\ldots, r,
$$
\be
 |\partial^r f^j(y)-\partial^r f^j(z)|\leq 
H\, ||y-z||^{\rho}, \; y,z \in \rr^d,\;\; j=1,2,\ldots, d \, \},
\label{2}
\ee
where $\partial^i f^j$ represents all partial derivatives of order $i$ 
of the $j$th component of $f$,
and $||\cdot||$ denotes the maximum norm in $\rr^d$. 
To assure that $f$ is a Lipschitz function, 
we assume that $\rho=1$ for $r=0$.
\f
We wish to compute a bounded function $l$ on $[a,b]$ that approximates 
the solution $z$. Letting $\{x_i\}$ be the uniform partition of $[a,b]$, 
$x_i=a+ih$ with $h=(b-a)/n$, the function $l$ will be produced 
by an algorithm $\phi$, based on approximations $a_i(f)$ 
to $z(x_i)$, $i=0,1,\ldots,n$.
\f
We now discuss the error and complexity models in the worst-case 
deterministic, randomized and quantum settings. 
In the worst-case deterministic setting,  
the error of $\phi$ at $f$ for the problem (\ref{1}) is defined by 
$$e(\phi,f)= \sup_{t\in [a,b]}  ||z(t) - l(t)||,$$
and the error in the class $F^{r,\rho}$ by
\be
e^{{\rm worst}}(\phi, F^{r,\rho}) = \sup_{f\in F^{r,\rho}} e(\phi,f) .
\label{3}
\ee
We assume that the values of
$f$ or its partial derivatives can be computed at given points
by a subroutine. The cost of an algorithm $\phi$ is measured by a 
number of subroutine calls. 
For a given $\e>0$, by the $\e$-complexity of the problem, 
$\comp^{{\rm worst}} (F^{r,\rho}, \e)$, we mean the minimal number of 
subroutine calls (taken among all possible algorithms)
sufficient to solve the problem with error at most $\e$, i.e., 
the minimal cost of an algorithm $\phi$ taken among all $\phi$
such that $e^{{\rm worst}}(\phi, F^{r,\rho}) \leq \e$.
\vsn
In the randomized setting, 
we allow a random selection of points at which the function $f$ 
is evaluated, so that the output of an algorithm is a random variable
(on a probability space ($\Omega$, $\Sigma$, $\P$)). 
Let the mappings $\omega\in \Omega \rightarrow a_i^\omega(f)$
be random variables for each $f\in F^{r,\rho}$
By an algorithm in the randomized  setting, we mean a tuple
\be 
\phi =( \{a_0^\omega(\cdot),a_1^\omega(\cdot), \ldots, a_n^\omega(\cdot)
\}_{\omega\in \Omega},\psi),
\label{3.1}
\ee
where $\psi$ is a mapping that produces a bounded function 
\be
l^\omega(t)=\psi(a_0^\omega(f), a_1^\omega(f),\ldots, 
a_n^\omega(f))(t)\, ,
\label{3.2}
\ee
$t\in [a,b]$, based on 
$a_0^\omega(f), a_1^\omega(f), \ldots, a_n^\omega(f)$. 
The error of $\phi$ at $f$ for the problem (\ref{1}) is defined by 
\be
e^\omega(\phi,f)=\sup_{t\in [a,b]}||z(t)-l^\omega(t)||.
\label{3.3}
\ee
We assume that the mapping
$\omega\in \Omega \rightarrow e^\omega(\phi,f)$
is a random variable with values in $\rr$, for each $f\in F^{r,\rho}$.
The error of $\phi$ in the class $F^{r,\rho}$ is
given by 
\be
e^{{\rm rand}}(\phi, F^{r,\rho}) = \sup\limits_{f\in F^{r,\rho}}
({\rm {\bf E}} e^\omega(\phi,f)^2)^{1/2} \, ,
\label{3a}
\ee
where {\bf E} is the expectation.
\f
As in the worst-case setting, we measure the cost of an algorithm $\phi$ by a 
number of subroutine calls that are needed to compute an approximation. 
For a given $\e>0$, by the $\e$-complexity of the problem, 
$\comp^{{\rm rand}} (F^{r,\rho}, \e)$, 
we mean the minimal cost of an algorithm $\phi$ taken among all $\phi$
such that  $e^{{\rm rand}}(\phi, F^{r,\rho}) \leq \e$.
\vsn
In the quantum setting, the output of an algorithm 
is a random variable (taking a finite number of values), but the reason 
of randomness is different than that in the randomized setting. 
On a quantum computer, where, roughly speaking,
basic objects are qubits (elements of  
a two-dimensional complex space $H_1$) and allowed 
operations are unitary transformations of the 
the tensor product of a number of copies of $H_1$,
randomness is a result of quantum measurement operations. 
For a detailed description of
the framework of numerical quantum computing one is referred to 
\cite{Heinrich}, where the notions of quantum measurement, quantum query,
quantum algorithm  and complexity  are defined and 
thoroughly  discussed, and applications to 
summation and integration problems are studied.
For a condensed discussion of randomized and quantum settings, 
in particular for the integration problem, 
one is also referred to \cite{HN}.
\f
By a quantum algorithm $\phi$ for solving our problem we mean a tuple 
(\ref{3.1}), 
where $a_i^\omega(f)$ are random approximations, 
in the quantum sense, to $z(x_i)$ for each $f$. 
The error of $\phi$ at $f$ is defined by (\ref{3.3}).
\f
Let $ 0<\delta<1/2$.
The error of $\phi$ in $F^{r,\rho}$ in the quantum setting 
is defined \cite{Heinrich} by
\be
e^{{\rm quant}}(\phi, F^{r,\rho},\delta) = \sup\limits_{f\in F^{r,\rho}}
\inf\; \{\; \alpha|\;\; \P\{\, e^{\omega}(\phi,f)>\alpha \,\}\; 
\leq \delta\; \}.
\label{3b}
\ee
Note that for a given $\e>0$ the bound $e^\omega(\phi,f)\leq \e$ 
holds with probability at least $1-\delta$ for each $f$ iff
$e^{{\rm quant}}(\phi, F^{r,\rho},\delta) \leq \e$.
\f
In the quantum setting the value of $\delta$ is usually set to $\delta=1/4$.
The error probability can then be reduced
to any $\delta$ by computing (componentwise) the 
median of $c\log 1/\delta$ repetitions of the algorithm, where $c$ 
is a positive number independent of $\delta$, 
see \cite{Heinrich1}, Lemma 3. For our problem, the
procedure of increasing the probability of success can be applied at
different levels which influences a logarithmic part of 
the cost of an algorithm, 
so that we shall describe it in more detail and discuss after the 
proof of Theorem~1. 
\f
On a quantum computer, the right-hand side function $f$ 
can be accessed through a query that
returns, for a  given point, a value of a component of $f$. 
Roughly speaking, a query on a class of real functions 
is defined as a transformation $Q$ that associates with 
each function $p$ a unitary mapping $Q_p$ defined on a Hilbert quantum space. 
For a detailed discussion of what is meant by "returning 
a value" of a function, and how 
a query is implemented in the quantum setting, the reader is 
referred to \cite{Heinrich} or \cite{Novak}. 
The~cost of an algorithm $\phi$ is measured by a 
number of quantum queries  that are needed to compute 
an approximation. (In upper bounds in Theorem 1, classical 
evalutions of $f$ or its partial derivatives are also taken into account.) 
For a given $\e>0$, by the quantum $\e$-complexity of the problem, 
$\comp^{{\rm quant}}(F^{r,\rho},\e,\delta)$, we mean 
the minimal cost of a quantum algorithm $\phi$ taken among all $\phi$
such that $e^{{\rm quant}}(\phi, F^{r,\rho},\delta) \leq \e$ . 
\f
We prove in this paper upper and lower bounds on the randomized and quantum 
complexity of initial-value problems (\ref{1}). The upper bounds,
summarized in the following theorem, will be next compared to
the known lower bounds on deterministic complexity to  
show that a speed-up is achieved in both settings. 
Lower bounds in the randomized and 
quantum settings are derived from a simple argument in the case 
$d\geq 2$, and are also included. We take below $\log=\log_2$
(although the base of the logarithm is not crucial).
\vsn
{\bf Theorem 1}$\;\;$ {\it For the problem (\ref{1}), we have that
\f
\be
\comp^{{\rm rand}} (F^{r,\rho}, \e) = 
O\left( \left(\frac{1}{\e}\right)^
{ \frac{r+\rho +3/2}{(r+\rho +1/2)(r+\rho+1)} } 
\log \frac{1}{\e} \right) \, , 
\label{7}
\ee
\be
 \comp^{{\rm quant}} (F^{r,\rho}, \e, \delta) = 
O\left( \left( \frac{1}{\e}\right) ^{\frac{r+\rho+2}{(r+\rho+1)^2}} 
(\log \frac{1}{\e} + \log \frac{1}{\delta}) \right) \, . 
\label{6}
\ee
For  $d\geq 2$,
\be
\comp^{{\rm rand}} (F^{r,\rho}, \e) = 
\Omega \left( \left(\frac{1}{\e}\right)^{\frac{1}{r +\rho +1/2} } \right) \, ,
\label{7a}
\ee
and, for $0< \delta\leq 1/4$, 
\be
\comp^{{\rm quant}} (F^{r,\rho}, \e, \delta) \geq 
\comp^{{\rm quant}} (F^{r,\rho}, \e, 1/4) = 
\Omega \left( \left(\frac{1}{\e}\right)^
{\frac{1}{r+\rho+1}} \right) \, .
\label{6a}
\ee
The constants in the "$O$" and "$\Omega$" notation 
only depend on the class $F^{r,\rho}$, and are independent 
of $\e$ and $\delta$.}~\qed
\vsn
Upper bounds  (\ref{7}) and (\ref{6}) will be derived by 
defining  suitable algorithms, while the lower bounds (\ref{7a}) and (\ref{6a})
are equal to those 
on the complexity of randomized or quantum computation of integrals 
of a function of one variable.
\f
Before giving the proof, we make some comments on these results. 
If the values of $f$ or its partial derivatives can only be accessed,
the deterministic worst-case complexity of the problem (\ref{1}) 
is of the order $\e^{-1/(r+\rho)}$, 
see Theorem 3 in the next section.  Since
\be
 \frac{r+\rho+2}{(r+\rho+1)^2}< \frac{r+\rho +3/2}{(r+\rho +1/2)(r+\rho+1)} 
  < \frac{1}{r+\rho} \, ,
\label{8}
\ee
both randomized and quantum computation yield an
improvement over the deterministic setting over the entire range of $r$ 
and $\rho$ (we neglect the logarithmic factors). For instance, 
if $r=0$ and $\rho =1$, the worst-case complexity in the deterministic 
setting is of the order $\e^{-1}$, in the randomized setting it is 
bounded from above by  $\e^{-5/6}$,  while on the quantum computer 
by $\e^{-3/4}$. 
\f
The lower bounds coincide with those for the integration problem, 
see (\ref{11}) and (\ref{10}). To see what the size of the gap between 
the bounds is, note that 
the reciprocal of the exponent in $1/\e$ in the quantum case  is such that
$$\lim\limits_{r\to \infty} 
\left( \frac{(r+\rho+1)^2}{ r+\rho+2} - (r+\rho) \right)=0\, ,$$
so that it behaves for large $r$ as $r+\rho$, while the lower bound 
depends on $r+\rho+1$. In the randomized setting, the reciprocal 
of the exponent in $1/\e$ behaves for large $r$ like $r+\rho$, 
while the lower bound depends on $r+\rho+1/2$.
\f
We now recall  results on randomized and quantum computation of integrals, 
as well as those 
on deterministic solution of initial-value problems.
\vsn
{\Large\section{Randomized and Quantum Computation of Integrals 
and Deterministic Solution of Initial-Value Problems 
}}
\noindent
Quantum complexity of integration has been first studied by Novak \cite{Novak}.
The problem is to approximate the integral
\be
I(g) =\int\limits_{[0,1]^s} \, g(x)\, dx 
\label{9}
\ee
for functions $g:[0,1]^s \to \rr$  from a H\"older class with 
$r\geq 0$, $0< \rho \leq 1$ 
\be
{\tilde F}^{r,\rho}= \{\, g\in C^r([0,1]^s) \mid \;\;
|g(y)| \leq  \tilde{D}_0, \, |\partial^r g(y)-\partial^r g(z)|\leq 
 \tilde{H}\, ||y-z||^{\rho}, \; y,z \in [0,1]^s  \, \},
\label{91}
\ee
where $\partial^r g$ represents all partial derivatives of order $r$ 
of $g$.
\f
The error of an algorithm $\phi$ at $g$ for the integration problem (\ref{9}) 
in the randomized and quantum settings is defined by
$$e^\omega(\phi,g) = |I(g) - A^\omega(g)|,$$
where $A^\omega(g)$ is the output of $\phi$ (in the worst-case setting 
the definition is the same, only the output is deterministic).
The other definitions of errors in the class of functions 
and complexity remain the same as for the problem (\ref{1}), 
with $F^{r,\rho}$ replaced by $\tilde{F}^{r,\rho}$.
\f
Based on the results on the computation of the mean of $n$ numbers given by
Brassard {\it et al} \cite{Bras} (upper bound), and Nayak and Wu \cite{Nayak}
(lower bound), Novak \cite{Novak} showed the~following 
result in the quantum setting. 
For the result in the randomized setting, see \cite{N88}, p. 62.
Let $\gamma =(r+\rho)/s$.
\vsn
{\bf Theorem 2 $\;$ (\cite{N88}, \cite{Novak})} 
$\;\;$ {\it  For the problem (\ref{9}) we have 
\be
\comp^{{\rm rand}} ({\tilde F}^{r,\rho}, \e) \asymp \e^{-1/(\gamma+1/2)}\, ,
\label{11}
\ee
\be
\comp^{{\rm quant}} ({\tilde F}^{r,\rho},\e, 1/4) \asymp \e^{-1/(\gamma+1)}.
\label{10}
\ee
} \qed
\f
An upper bound in (\ref{11}) can be achieved by
random algorithms with a finite number of output values. 
\f
Consider now the solution of initial-value problem (\ref{1}) 
in the deterministic setting. This problem has been considered in 
a number of papers, see, e.g., \cite{Kac1}  or \cite{Kac2}. 
The following result is 
a straightforward modification of Corollary 4.1 from \cite{Kac1}. 
The modification is needed, since  the class of functions 
considered in \cite{Kac1}, consisting of $r$ times continuously
differentiable functions with bounded derivatives, is to be replaced 
with the H\"older class $F^{r,\rho}$. For the modification in lower bounds,
one is referred to 
the proof of Theorem 3.1 from \cite{Kac2}, where the functions $g_k$
in the construction must be replaced by suitable 
functions from the class 
$F^{r,\rho}$. The upper bounds will be derived again in the sequel, 
as a by-product in the proof of Theorem 1.
\f
{\bf Theorem 3 $\;$ (\cite{Kac1}) } $\;\;$ {\it In the deterministic setting, 
the complexity of (\ref{1}) satisfies: 
\vsn
-- if the values of $f$ or its partial derivatives 
are only accessible, then
\be
\comp^{{\rm worst}} ( F^{r,\rho}, \e) \asymp \e^{-1/(r+\rho)}\, ,
\label{12}
\ee
-- if arbitrary linear functionals are accessible, then
\be
\comp^{{\rm worst}} ( F^{r,\rho}, \e) \asymp \e^{-1/(r+\rho +1)}\, .
\label{13}
\ee
} \qed
\f
(The lower bound in (\ref{13}) holds true not only for linear functionals, but also 
for a class of nonlinear functionals, see \cite{Kac1}.) 
Relation (\ref{13}) will play an important role in the proof of Theorem~1, 
while (\ref{12})  will serve as a point of reference  
in evaluating a speed-up obtained due to randomization or due to
quantum computations.
\f
Let us now recall 
the algorithm that leads to the upper bound in (\ref{13}). It requires 
the computation of integrals of $f$, and is defined as follows.
\vsn 
Take $y_0^* =\eta$. Given $y_i^*\;$ ($y_i^* \cong z(x_i)$), 
we let $\bar{z}_i^*(t)$ be the solution of the problem
\be
\bar{z}'(t) = f(\bar{z}(t)), \;\;\; t\in [x_i,x_{i+1}], \;\;\;\;\;
\bar{z}(x_i)=y_i^* ,
\label{14}
\ee
and we set 
\be
l_i^*(t)= \sum\limits_{j=0}^{r+1} \, \frac{1}{j!} 
\bar{z}_i^{*\, (j)}(x_i)\, (t-x_i)^j\, ,
\;\;\;\; t\in [x_i,x_{i+1}]\, .
\label{15}
\ee
Then we define 
\be
y_{i+1}^*=y_i^* +\int\limits_{x_i}^{x_{i+1}}\, f(l_i^*(t))\, dt\, ,
\label{16}
\ee
$i=0,1,\ldots, n-1,$ and finally
\be
l(t)=l^*_i(t) \;\;\; \mbox{ for } t\in [x_i,x_{i+1}].
\label{161}
\ee
(The function $l$ is piecewise continuous. It is also possible to define it 
to be continuous on $[a,b]$.)
\vsn
{\Large \section{Randomized and Quantum 
Solution of Initial-Value Problems}}
\noindent
We now define randomized and quantum algorithms for 
the solution of~(\ref{1}). Let $w_i^*$ be a polynomial
\be
w_i^*(y)=\sum\limits_{j=0}^r \, \frac{1}{j!} f^{(j)}(y_i^*)(y-y_i^*)^j \, ,
\label{17}
\ee
where $f^{(j)}(y_i^*)z^j$ is meant to be the
value of the $j$-linear operator $f^{(j)}(y_i^*)$
at $(z,z\ldots,z)\,$ ($j$ times). The 
values of $w_i^*$ can be computed through evaluation of partial 
derivatives of components of $f$ of order $0,1,\ldots, r$.
Equality (\ref{16}) can be equivalently written as 
\be
y_{i+1}^*=y_i^* +\int\limits_{x_i}^{x_{i+1}}\, w^*_i(l_i^*(t)) \,dt
+ h^{r+\rho+1} \int\limits_{0}^{1}\, g_i(u)\, du \, ,
\label{19}
\ee
where
\be
g_i(u)= \frac{1}{h^{r+\rho}} \left( f(l_i^*(x_i+uh))- w^*_i(l_i^*(x_i+uh)) 
\right)\, , 
\label{20}
\ee
for $u\in [0,1]$ and $i=0,1,\ldots, n-1\,$ .
One can verify that $g_i$ belongs to $C^{(r)}([0,1])$, 
the derivatives of $g_i$ 
are bounded by constants that depend only on  the parameters of the 
class $F^{r,\rho}$ (and are independent of $i$, $y_i^*$ and $h$), and 
$$||g_i^{(r)}(u)-g_i^{(r)}(\bar{u})|| \leq {\tilde H}\, |u-\bar{u}|^{\rho},$$
$u, \bar{u} \in [0,1]$, for some constant ${\tilde H}$ depending on 
the parameters as above.
\f
The algorithm (deterministic, randomized or quantum) 
for solving (\ref{1}) is defined as follows (we omit the argument $\omega$
in random variables). 
Let $a_0(f)=y_0=\eta$. Given $a_i(f)=y_i$, we consider 
functions $g_i$ defined by (\ref{20}) for $y_i$ (that is,  
the polynomials $l_i^*$ and $w_i^*$ based on $y_i^*$
are replaced by the polynomials $l_i$ and $w_i$ based on $y_i$), 
and compute some approximations $A_i(f)$ 
to the integrals $\int\limits_0^1 \, g_i(u)\, du$ in (\ref{19}). 
The algorithm is defined by setting $a_{i+1}(f)=y_{i+1}$, where 
\be
y_{i+1}=y_i +\int\limits_{x_i}^{x_{i+1}}\, w_i(l_i(t))\, dt
+ h^{r+\rho+1} A_i(f) \, , 
\label{23}
\ee
and the approximation on $[x_i,x_{i+1}]$ is given by
\be
l(t) = l_i(t) \, ,
\label{23a}
\ee
$i=0,1,\ldots, n-1$. Finally, we set $\; \psi(a_0(f),\ldots,a_n(f))(t)=l(t)$ 
for $t\in [a,b]$. 
\f
The approximations $A_i(f)$ may be obtained by  
deterministic, randomized or quantum  algorithms. 
In  the randomized and quantum settings, we demand random variables 
$A_i(f)$ to satisfy
\be
\left|\left|\int\limits_0^1 \, g_i(u)\, du - A_i (f)\right|\right|\leq \e_1
\;\;\; \mbox{ with probability  at least } (1-\delta)^{1/n}
\label{231}
\ee
for $i=0,1,\ldots, n-1$ (and all $y_i$), for some $\e_1$.
It will be shown later on that a satisfactory choice is $\e_1=h$.
\vsn
For illustration, we specify the algorithm above in the case 
$r=0$. It may be considered as a modification of
Euler's method and is defined as follows. Given $y_i$, we compute 
$$
y_{i+1}=y_i + hf(y_i) + h^{1+\rho}A_i(f),
$$
where $A_i(f)$ is a (deterministic, randomized or quantum)
approximation to $(1/h^{\rho})\int\limits_0^1 
(f(y_i +uhf(y_i))-f(y_i))\, du\; $ 
with error at most $\e_1$, and probability at least $(1-\delta)^{1/n}$ 
(in non-deterministic cases), $i=0,1,\ldots, n-1$. The approximation 
to $z=z(t)$ is defined  on $[x_i, x_{i+1}]$ by $l(t)=y_i + f(y_i)(t-x_i)$.
\vsn
{\bf Proof of Theorem 1}
\vsn
Consider first the quantum setting. We use the quantum algorithm of Novak  
to compute $A_i(f)$ (componentwise in a statistically independent way). 
Due to (\ref{10})  with $s=1$, we have that for each $i$ the inequality
\be
\left|\left|\int\limits_0^1 \, g_i(u)\, du - A_i (f)\right|\right|\leq \e_1
\label{232}
\ee
holds with probability  at least  $3/4$, and the 
quantum query cost $O(\e_1^{-1/(r+\rho +1)})$.
(Since $g_i$ has $d$ components, only the constant in the $"O"$ 
notation is different than that in (\ref{10}), which is a result of 
computing the median of a suitable number of 
repetitions to increase the probability of
success in each component to $(3/4)^{1/d}$.)
Taking componentwise the median of $k$ repetitions, 
 $$k =\Theta \left(\log \frac{1}{1-(1-\delta)^{1/n}} \right) \, ,$$
 we arrive at an approximation $A_i (f)$ (the same symbol is used 
to denote this new approximation) such that (\ref{231}) is satisfied. 
 Since 
 $$\log \frac{1}{1-(1-\delta)^{1/n}} \leq c(\log n + \log 1/\delta)$$
 (where $c$ is independent of $n$ and $\delta$), 
the  cost of computing $A_i(f)$  is of order 
$O(\e_1^{-1/(r+\rho +1)}\, (\log n + \log 1/\delta)$. 
Thus, with the cost $O(\e_1^{-1/(r+\rho +1)} \, n\,(\log n + \log 1/\delta))$ 
we assure that the bounds 
\be
\left|\left|\int\limits_0^1 \, g_i(u)\, du - A_i (f)\right|\right|
\leq \e_1 \;\;\;
\mbox{ for } i=0,1,\ldots, n-1
\label{233}
\ee
hold simultanously with probability at least $1-\delta$. 
The total cost of the 
algorithm additionally includes $O(n)$ classical subroutine calls necessary
in the deterministic part of the algorithm. 
\f
Let $e_i=z(x_i) - y_i$. Since the solution of (\ref{1}) satisfies 
\be
z(x_{i+1}) = z(x_i) + \int\limits_{x_i}^{x_{i+1}}  f(z(t)) \, dt\, ,
\label{24}
\ee
 we get from (\ref{23}) that
\be
\begin{array}{ll}
e_{i+1} = e_i & + \int\limits_{x_i}^{x_{i+1}} 
( f(z(t)) - f(l_i(t)) ) \, dt  \\
& + \int\limits_{x_i}^{x_{i+1}} ( f(l_i(t)) - w_i (l_i(t)) )\, dt - 
h^{r+\rho +1} A_i(f) \, .
\end{array}
\label{26}
\ee
We shall derive from (\ref{26}) a difference inequality for $||e_i||$.  
Let 
$\bar{z}_i$ denote the solution of the local problem (\ref{14})
with the initial condition $\bar{z}_i(x_i) = y_i$. 
By triangle inequality,
\be
|| f(z(t)) - f(l_i(t)) ||\leq || f(z(t)) - f(\bar{z}_i(t))|| +
|| f(\bar{z}_i(t)) - f(l_i(t)) ||.
\label{t}
\ee
To estimate the first term, we note that 
the dependence of the solution on initial condition yields that
\be
||z(t) - \bar{z}_i(t) || \leq \exp(Lh) ||z(x_i) - y_i||, \;\;\;\;
t\in [x_i,x_{i+1}],
\label{261}
\ee
where $L$ is the Lipschitz constant of $f$.
Writing the remainder of Taylor's formula in the form  
$$
\bar{z}_i(t)-l_i(t) = \int\limits_0^1 
(\bar{z}_i^{(r+1)} (\theta t+(1-\theta) x_i) - \bar{z}_i^{(r+1)} (x_i) )
(t-x_i)^{r+1} (1-\theta)^r/r!\, d\theta\, ,
$$
$t\in [x_i,x_{i+1}]$, and checking that $\bar{z}_i^{(r+1)}$ 
is a H\"older function
with exponent $\rho$  (and a constant that only depends on
the parameters of the class $F^{r,\rho}$ and is 
independent of $i, y_i, n$), we arrive at 
\be
 ||\bar{z}_i (t)-l_i(t)||\leq M\, h^{r+\rho+1}\, , 
 \label{2611}
\ee
$t\in [x_i,x_{i+1}]$, where the constant $M$ only depends on 
the parameters of the class $F^{r,\rho}$. 
\f
The last two inequalities together with (\ref{26}), (\ref{t}) and (\ref{233})
yield the relation (satisfied with probability at least $1-\delta$)
\be
||e_{i+1}||\leq ||e_i|| \left(1+hL\exp(Lh)\right) + LM h^{r+\rho+2} +
h^{r+\rho+1}\e_1 \, ,
\label{27}
\ee
$i=0,1,\ldots,n-1$. 
Take now $\e_1=h$. By solving difference inequality (\ref{27}), we get that
\be
\max\limits_{0\leq i\leq n} ||e_i|| = O( h^{r+\rho +1})
\label{28}
\ee
with probability at least $1-\delta$. Finally, for any 
$t\in [x_i,x_{i+1}]$ we have 
due to (\ref{261}), (\ref{2611}) and  (\ref{28})  that 
\be
||z(t)-l(t)||=||z(t)-l_i(t)||\leq ||z(t)-\bar{z}_i(t)||+||\bar{z}_i(t)
-l_i(t)||= O(h^{r+\rho+1}),
\label{281}
\ee
with the constant depending only on the parameters 
of the class $F^{r,\rho}$, and probability at least $1-\delta$. 
\f
The quantum query cost  of the considered algorithm is 
of order $O( n^{(r+\rho+2)/(r+\rho +1)} (\log n + \log 1/\delta) )$.
For $\e>0$, we now take the minimal $n$ such that the upper bound 
in (\ref{281}) does not exceed $\e$, $\, n \asymp (1/\e)^{1/(r+\rho +1)}\, $, 
and the desired upper bound on the complexity (\ref{6}) follows.
\f
In the randomized setting we use (\ref{11}). Take an algorithm $\phi$
for approximating integrals with $s=1$ such that
$e^{{\rm rand}}(\phi,\tilde{F}^{r,\rho}) \leq \e_1/2$. By the Chebyshev 
inequality we have that
$$ \P \left\{ \left| \int\limits_0^1 g(x)\, dx -  A^\omega(g)\right| > \e_1
\right\}
\leq 1/4, $$
for any $g\in \tilde{F}^{r,\rho}$,
where $A^\omega(g)$ is the output of the algorithm.
 As in the quantum case, the inequality 
(\ref{232}) holds with probability  at least  $3/4$, but the 
cost is now $O(\e_1^{-1/(r+\rho +1/2)})$. 
Proceeding further on similarly as in the quantum case, 
one gets that  the bound
(\ref{281}) holds with probability at least $1-\delta$, 
and cost 
$O(n^{(r+\rho +3/2)/(r+\rho+1/2)}(\log n + \log 1/\delta))$. 
We denote the resulting random algorithm for initial value 
problems by $\tilde{\phi}$. 
\f
Take now $n$ to be the minimal number for which the upper bound
in (\ref{281}) is at most $\e/2$, $\, n\asymp (1/\e)^{1/(r+\rho +1)}\, $, 
and observe that
$$\E e^\omega(\tilde{\phi}, f)^2 =\int\limits_{e^{\omega}(\tilde{\phi},f)>
\e/2} e^{\omega}(\tilde{\phi},f)^2 \, d \P(\omega)
+  \int\limits_{e^{\omega}(\tilde{\phi},f)\leq 
\e/2} e^{\omega}(\tilde{\phi},f)^2 \, d \P(\omega) 
\leq K^2\delta + \e^2/4$$
for all $f\in F^{r,\rho}$, where $K$ is a positive constant that depends only
on the parameters of the class $F^{r,\rho}$ such that 
$e^\omega(\tilde{\phi}, f)\leq K$ (such a constant exists, 
since we may assume that $|A^{\omega}(g)|\leq 2\tilde{D}_0$;
otherwise $A^{\omega}(g)=0$ would be a better approximation).
Hence, the algorithm $\tilde{\phi}$ with $\delta = 3\e^2/4K^2$ 
satisfies $e^{{\rm rand}} (\tilde{\phi}, F^{r,\rho})\leq \e$. 
Looking at the cost of $\tilde{\phi}$ we see that 
the upper bound (\ref{7}) is proven.
\f
Consider now the deterministic setting. 
The known upper bounds on the complexity can 
be derived again as follows. If the values of $f$
or its partial derivatives can only be accessed, then we are able to   
approximate $\int\limits_0^1 \, g_i(u)\, du$ within the error $\e_1$
with cost $O(\e_1^{-1/(r+\rho)})$. Since $n$ integrals are to be approximated,
the total cost of the algorithm (for $\e_1=h$),   
with the error bound (\ref{28}), is of order 
$O(n\cdot n^{1/(r+\rho)})$. This proves the upper bound in (\ref{12}).
If exact computation of the integrals is allowed in the model, 
then we simply take $\e_1=0$, 
which leads to the upper bound in (\ref{13}).
\vsn
We now pass to lower bounds. Consider the quantum setting. 
Let $g\in \tilde{F}^{r,\rho}$ with $r\geq 1$ and $\rho\in (0,1]$,
or $r=0$ and $\rho =1$, and with $s=1$.
Consider a two-dimensional problem
\be
\left\{ 
\begin{array}{lll}
u'(t)&=&1\\
v'(t)&=&g(u(t)) \; ,\;\;\;\; t\in [0,1]\, ,
\end{array} 
\right.
\label{29}
\ee
with initial conditions $u(0)=0$, $v(0)=0$. 
The function $g$ can be extended to $\rr$ such that 
the right-hand side function in (\ref{29}) 
belongs to the class $F^{r,\rho}$ with suitably chosen parameters. 
The solution of (\ref{29}) 
is given by $u(t)=t$ and $v(t)=\int\limits_0^t g(s)\, ds$. 
\f
Let $\e>0$ and let $\phi$ be any quantum algorithm for solving (\ref{1}) 
with error 
$e^{{\rm quant}}(\phi,F^{r,\rho},1/4) \leq \e$, and quantum query cost $c(\e)$. 
When applied to (\ref{29}), 
the algorithm $\phi$ gives an approximation to 
$v(1)=\int\limits_0^1\, g(t)\, dt$ with error at most $\e$, 
with probability at least $3/4$.  
Due to the lower bound for integration in (\ref{10}) the cost must be
at least of order $\e^{-1/(r+\rho+1)}$ queries on $g$, which yields that 
$c(\e) = \Omega(\e^{-1/(r+\rho+1)})$, and proves lower bound (\ref{6a}). 
\f
In the randomized setting we use similar arguments to show (\ref{7a}),  
adjusted to the error formula (\ref{3a}). 
Since the same arguments apply in both settings for any $d\geq 2$, 
the proof of Theorem 1 is completed. \qed
\vsn
Let us finally note that the logarithmic factor in the upper bound
(\ref{6}) depends on what stage the median is computed at in 
the algorithm. If we  ask 
the inequality (\ref{231}) to hold with probability at least $(3/4)^{1/n}$,
proceed up to the final step (\ref{281}) with $\delta=1/4$, and after that
compute the median of $c\log 1/\delta$ repetitions of the entire algorithm, 
then the factor $\log 1/\epsilon+\log 1/\delta$ in (\ref{6}) would be 
replaced by $\log 1/\epsilon\cdot\log 1/\delta$. 
\vsn

\end{document}